\documentclass[twocolumn,tighten]{aastex6}
\usepackage{amsmath,amstext}
\usepackage[T1]{fontenc}
\usepackage{apjfonts}
\usepackage[figure,figure*]{hypcap}

\shorttitle{Apparently decaying orbit of WASP-12b}
\shortauthors{Patra et al.}

\begin{document}

\title{The Apparently Decaying Orbit of WASP-12\lowercase{b}}

\author{Kishore C.~Patra\altaffilmark{1}}
\author{Joshua N.~Winn\altaffilmark{2}}
\author{Matthew J.~Holman\altaffilmark{3}}
\author{Liang Yu\altaffilmark{1}}
\author{Drake Deming\altaffilmark{4}}
\author{Fei Dai\altaffilmark{1,2}}

\altaffiltext{1}{Department of Physics, and Kavli Institute for
  Astrophysics and Space Research, Massachusetts Institute of
  Technology, Cambridge, MA 02139, USA}

\altaffiltext{2}{Department of Astrophysical Sciences, Princeton
  University, 4 Ivy Lane, Princeton, NJ 08540, USA}

\altaffiltext{3}{Harvard-Smithsonian Center for Astrophysics, 60
  Garden Street, Cambridge, MA 02138, USA}

\altaffiltext{4}{Department of Astronomy, University of Maryland at
  College Park, College Park, MD 20742, USA}

\begin{abstract}
  We present new transit and occultation times for the hot Jupiter
  WASP-12b. The data are compatible with a constant period derivative:
  $\dot{P}=-29 \pm 3$~ms~yr$^{-1}$ and $P/\dot{P}= 3.2$~Myr. However,
  it is difficult to tell whether we have observed orbital decay or a
  portion of a 14-year apsidal precession cycle. If interpreted as
  decay, the star's tidal quality parameter $Q_\star$ is about
  $2\times 10^5$. If interpreted as precession, the planet's Love
  number is $0.44\pm 0.10$. Orbital decay appears to be the more
  parsimonious model: it is favored by $\Delta\chi^2=5.5$ despite
  having two fewer free parameters than the precession model. The
  decay model implies that WASP-12 was discovered within the final
  $\sim$0.2\% of its existence, which is an unlikely coincidence but
  harmonizes with independent evidence that the planet is nearing
  disruption. Precession does not invoke any temporal coincidence, but
  it does require some mechanism to maintain an eccentricity of
  $\approx$0.002 in the face of rapid tidal circularization. To
  distinguish unequivocally between decay and precession will probably
  require a few more years of monitoring. Particularly helpful will be
  occultation timing in 2019 and thereafter.
\end{abstract}

\keywords{planets and satellites: individual (WASP-12 b) --- planet-star interactions}

\section{Introduction}
\label{sec:intro}

More than 20 years have elapsed since the discovery of hot Jupiters
\citep{MayorQueloz1995}. The time may be ripe to confirm a
long-standing theoretical prediction: the orbits of almost all of these
planets should be shrinking due to tidal orbital decay
\citep{Rasio+1996, Sasselov2003, Levrard+2009}.  This is because the
star's rotational angular momentum is typically smaller than one-third
of the orbital angular momentum, the critical value beneath which
tidal evolution has no stable equilibrium \citep{Hut1980}.

Tidal decay of hot Jupiters has been invoked to explain certain
properties of the ensemble of star-planet systems. For example, the
scarcity of gas giants with periods less than a day is suggestive of
orbital decay \citep[see,
  e.g.][]{Jackson+2008,Hansen2010,Penev+2012,Ogilvie2014}. The
anomalously rapid rotation of some hot-Jupiter host stars has been
attributed to transfer of the planet's orbital angular momentum
\citep{Penev+2016}. The absence of hot Jupiters around subgiant stars
may be caused by an acceleration of orbital decay when a star leaves
the main sequence
\citep{VillaverLivio2009,Hansen2010,SchlaufmanWinn2013}.  Tidal decay
might also be responsible for the lower occurrence of close-in planets
around rapidly rotating stars \citep{TeitlerKonigl2014}, or the
realignment of stars and their planetary orbits
\citep{MatsakosKonigl2015}. However, direct evidence for orbital decay
has been lacking: there have been no clear demonstrations of a
long-term period decrease due to orbital decay \citep[see,
  e.g.,][]{Hoyer+2016,Wilkins+2017}.

Another unfulfilled prediction is that the orbits of hot Jupiters
should be apsidally precessing on a timescale of decades
\citep{MiraldaEscude2002,HeylGladman2007,PalKocsis2008,JordanBakos2008},
as long as the orbits are at least slightly eccentric. In particular,
\citet{RagozzineWolf2009} noted that the theoretical precession rate
is dominated by the contribution from the planet's tidally deformed
mass distribution. They advocated a search for apsidal precession as
a means of probing the interiors of hot Jupiters.

With an orbital period of 1.09 days, WASP-12b is one of the
shortest-period giant planets known \citep{Hebb+2009}, and has been
monitored for a decade.  It is, therefore, an outstanding target in the
search for orbital decay and apsidal precession.
\citet{Maciejewski+2016} reported a decrease in the apparent
period. Despite being the most convincing claim that has yet been
presented for orbital decay, those authors could not distinguish
between true period shrinkage and a long-term oscillation of the
apparent period due to apsidal precession. In this paper, we present
new transit and occultation times (\S~\ref{sec:transits}
and~\ref{sec:occultations}). We use all of the available data to test
which model is favored by the data: a constant period derivative, or
sinusoidal variations arising from apsidal precession
(\S~\ref{sec:timing}). We also discuss the implications of both models
(\S~\ref{sec:implications}) and prospects for future observations
(\S~\ref{sec:future}).

\section{New transit times}
\label{sec:transits}

\begin{figure}[ht]
    \epsscale{1.0}
    \plotone{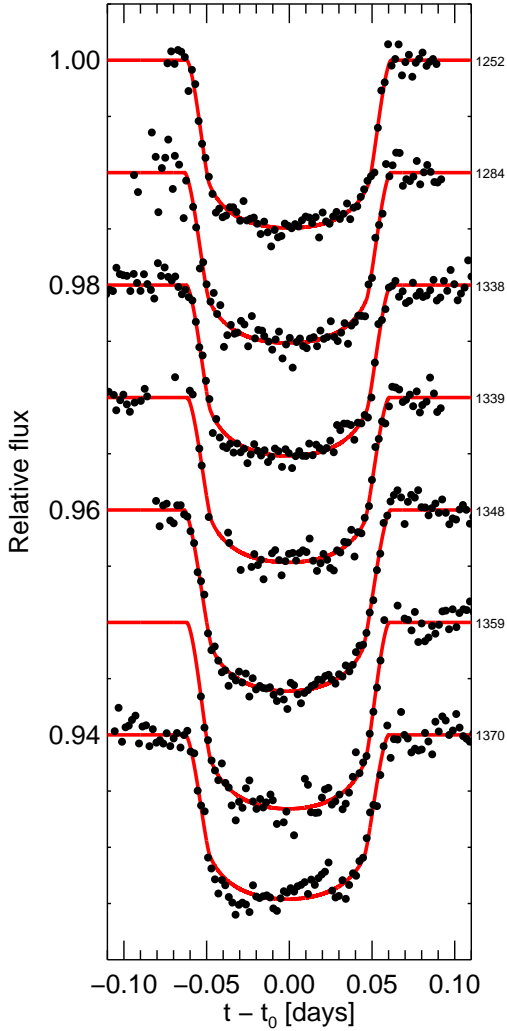}
    \caption{{\bf New transit light curves.} Black points are based on
      observations with the FLWO~1.2m telescope in the Sloan $r'$
      band. Red curves are the best-fit models. Epoch numbers are
      printed to the right of each curve. Vertical offsets have been
      applied to separate the light curves.\label{fig:lightcurves}}
\end{figure}

Between 2016~October and 2017~February, we observed seven transits of
WASP-12 with the 1.2m telescope at the Fred Lawrence Whipple
Observatory on Mt.\ Hopkins, Arizona.  Images were obtained with the
KeplerCam detector through a Sloan $r'$-band filter. The typical
  exposure time was 15~s, chosen to give a signal-to-noise ratio
  of about 200 for WASP-12. The field of view of this camera is
$23\farcm 1$ on a side.  We used 2x2 binning, giving a pixel scale of
$0\farcs 68$.

The raw images were processed by performing standard overscan
correction, debiasing, and flat-fielding with IRAF\footnote{The Image
  Reduction and Analysis Facility (IRAF) is distributed by the
  National Optical Astronomy Observatory, which is operated by the
  Association of Universities for Research in Astronomy (AURA) under a
  cooperative agreement with the National Science
  Foundation.}. Aperture photometry was performed for WASP-12 and an
ensemble of 7-9 comparison stars of similar brightness.  The
  aperture radius was chosen to give the smallest scatter in the flux
  outside of the transits, and was generally 7-8 pixels.  The
reference signal was generated by summing the flux of the comparison
stars. The flux of WASP-12 was then divided by this reference signal
to produce a time series of relative flux.  Each time series was
normalized to have unit flux outside of the transit.  The time stamps
were placed on the BJD\textsubscript{TDB} system using the code of
\citet{Eastman2010}.

We fitted a \citet{MandelAgol2002} model to the data from each
transit. The parameters of the transit model were the midtransit time,
the planet-to-star radius ratio ($R_{\rm p}/R_\star$), the scaled
stellar radius ($R_\star/a$), and the impact parameter ($b = a\cos
i/R_\star$). For given values of $R_\star/a$ and $b$, the transit
timescale is proportional to the orbital period [see, e.g., Eqn.~(19)
  of \citet{Winn2010}]. To set this timescale, we held the period fixed
at $1.09142$~days, although the individual transits were fitted
separately with no requirement for periodicity.  To correct for
differential extinction, we allowed the apparent magnitude to be a
linear function of airmass, giving two additional parameters. The limb
darkening law was assumed to be quadratic, with coefficients held
fixed at the values ($u_1 = 0.32$, $u_2= 0.32$) tabulated by
\citet{ClaretBloemen2011} for a star with the spectroscopic parameters
given by \citet{Hebb+2009}.\footnote{For this purpose we used the online code of
  \citet{Eastman+2013}:
  http://astroutils.astronomy.ohio-state.edu/exofast/limbdark.shtml}

To determine the credible intervals for the parameters, we used the
{\it emcee} Markov Chain Monte Carlo (MCMC) code written by
\citet{ForemanMackey+2013}. The transition distribution was proportional
to $\exp(-\chi^2/2)$ with
\begin{equation}
\chi^2 = \sum_{i=1}^{N}\left(\frac{ f_{{\rm obs}, i} - f_{{\rm calc}, i}}{\sigma_i}\right)^2,
\end{equation}
where $f_{{\rm obs}, i}$ is the observed flux at time $t_i$ and
$f_{{\rm calc}, i}$ is the corresponding flux of the model.  The
uncertainties $\sigma_{i}$ were set equal to the standard deviation of
the out-of-transit data.
In a few cases, the pre-ingress scatter was
noticeably different than the post-egress scatter; for those
observations, we assigned $\sigma_i$ by linear interpolation between
the pre-ingress and post-egress values.

Figure~\ref{fig:lightcurves} shows the light curves and the
best-fit models. Table~1 reports the midtransit times and their
uncertainties. For convenience, this table also includes the new
occultation times described below, as well as the previously reported
times that are analyzed in Section~\ref{sec:timing}. The results for
the other transit parameters were consistent with the previous results
of \citet{Maciejewski+2013}, with larger uncertainties.

Time-correlated noise is evident in some of the new light
curves. Although we made no special allowance for these correlations
in our analysis, we have reason to believe that the quoted
uncertainties are reliable.  When these seven new midtransit times are
fitted with a linear function of epoch, we obtain $\chi^2_{\rm
  min}=5.1$ with five degrees of freedom.  When the period is held
fixed at the value derived from all 10 years of timing data, we obtain
$\chi^2_{\rm min}=7.8$ with six degrees of freedom. These tests
suggest that the uncertainties are not substantially
underestimated. Furthermore, spurious timing variations would be
random from night to night, whereas our long-term timing analysis
(Section~\ref{sec:timing}) reveals that all seven new midtransit times
produce residuals of the same sign and amplitude.

\begin{figure*}[ht!]
  \begin{center}
    \leavevmode
\includegraphics[width=0.9\linewidth]{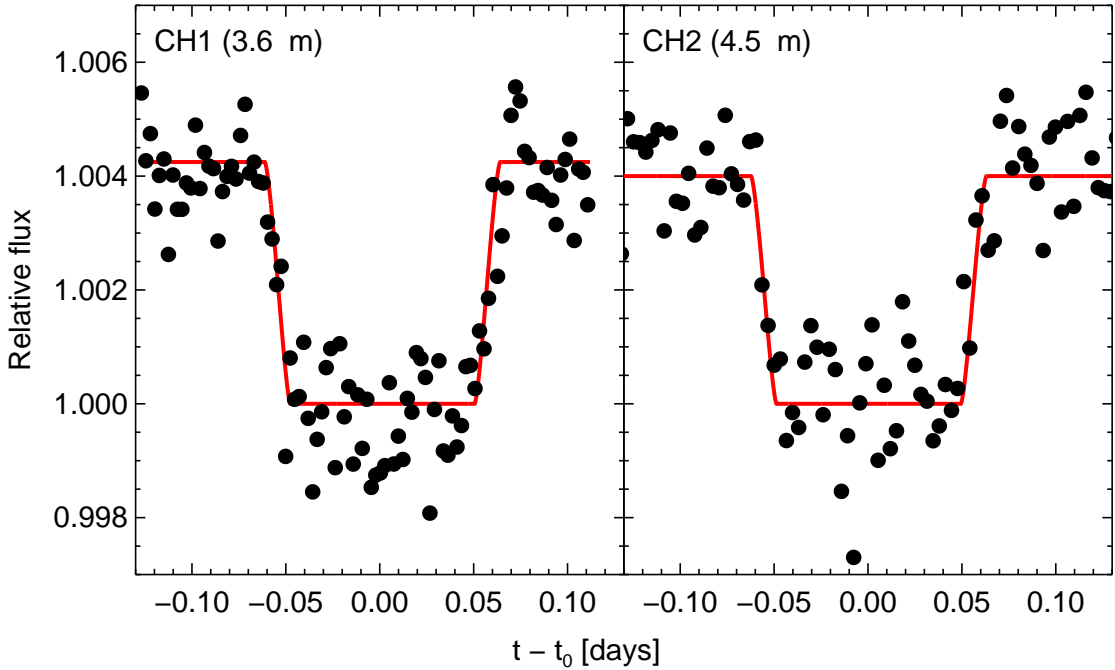}
  \end{center}
  \caption{{\bf New occultation light curves.} Black points are the
    binned {\it Spitzer} measurements from epochs 305 (left) and 308
    (right). Red curves are the best-fit
    models. \label{fig:occultations}}
\end{figure*}

\newpage

\section{New occultation times}
\label{sec:occultations}

We measured two new occultation times based on hitherto unpublished
{\it Spitzer} observations in 2013 December (program 90186,
P.I.~Todorov).  Two different transits were observed, one at
  3.6~$\mu$m and one at 4.5~$\mu$m.  The data take the form of a time
series of 32x32-pixel subarray images, with an exposure time of 
  2.0~s per image. The data were acquired over a wide range of
orbital phases, but for our purpose, we analyzed only the
$\approx$14,000 images within 4 hours of each occultation. We also
reanalyzed the {\it Spitzer} occultation presented by
\citet{Deming+2015} using the technique described below.

We determined the background level in each image by fitting a Gaussian
function to the histogram of pixel values, after excluding the high
flux values associated with the star. The centroid of the fitted
Gaussian function was taken to be the background value and was
subtracted from each image prior to performing aperture photometry.

We used two different schemes to choose photometric aperture sizes.
In the first scheme, we used 11 apertures ranging in radius from
1.6-3.5 pixels in average increments of 0.2~pixel. In the second
scheme, we tried 11 apertures for which the radius was allowed to vary
at each time step, based on the procedure described in Appendix A of
\citet{Lewis+2013}.  In this procedure, the aperture radius is
taken to be the sum of a constant (ranging from 0-2 pixels) and the noise
pixel radius, defined as the square root of the ratio of the
square of the total flux integrated over all pixels divided by the sum
of the squared-fluxes in individual pixels.  The noise pixel radius is
specific to each image and allows for possible changes in the shape of
the pixel response function with position. We also
tried two different methods to choose the center of the apertures:
fitting a two-dimensional Gaussian function to the stellar image, and computing
the flux-weighted center-of-light. Hence, there were four versions of
the photometry: constant versus variable aperture radii, and Gaussian
centroiding versus center-of-light. Each of those four versions
contains 11 time series with different aperture sizes.

We corrected for the well-known intrapixel sensitivity variations
using pixel-level decorrelation [PLD; \citet{Deming+2015}]. In PLD,
the flux time series is modeled as the sum of the astrophysical
variation, a temporal baseline, and a weighted sum of the (normalized)
time series of each pixel comprising the point-spread function.
Because each pixel value is divided by the total brightness of the
star in that image, PLD effectively separates astrophysical
information and {\it Spitzer} detector effects. PLD has also been used
to produce high-quality photometry from {\it K2} data
\citep{Luger+2016}.

Our implementation of PLD operates on time-binned data [see Sec.~3.1
  of \citet{Deming+2015}]. Over a trial range of occultation midpoints
and median aperture radii, the code uses linear regression to find the
best-fit occultation depth and pixel coefficients. We
provisionally adopt the midpoint that produces the best fit (smallest
$\chi^2$). The code then varies the aperture radius from among the 11
possible values and the duration of the time bins. The optimal values
of the radius and bin size are determined by examining the
\citet{Allan1966} deviation relation of the residuals and identifying
the case that comes closest to the ideal relation.\footnote{The Allan
  deviation relation expresses how the standard deviation of the
  binned residuals varies with bin size. For ideal white noise, it
  should decrease as the inverse square root of the bin size.} Then, an
MCMC procedure is used to optimize the light-curve parameters (including
the time of mid-occultation), pixel coefficients, and temporal
baseline coefficients. The temporal baseline was taken to be a
quadratic function of time, which was sufficient to describe the
phase-curve variation in the vicinity of the occultation.

After performing these steps for all four different versions of the
photometry, we adopted the version that came closest to achieving the
theoretical photon noise limit. For the 3.6\,$\mu$m data, the adopted
version used 10-frame binning, center-of-light centroiding, and a
constant aperture radius of 2.3 pixels. For the 4.5\,$\mu$m data, the
adopted version used 10-frame binning, center-of-light centroiding,
and a constant aperture radius of 2.2 pixels. With these choices, we
achieved a noise level of 1.29 and 1.24 times the theoretical photon
noise limit at 3.6 and 4.5\,$\mu$m, respectively. The uncertainty in
the midpoint of each occultation was determined from the standard
deviation of the (very nearly Gaussian) marginalized posterior
distribution. The new light curves are shown in
Figure~\ref{fig:occultations}, and the times are given in Table~1. The
best-fit central times are relatively insensitive to the version
of the photometry adopted in the final solution.  The very worst of
the four photometry solutions for the 3.6 and 4.5\,$\mu$m data gave
midpoints differing by 31 and 75 seconds (0.3$\sigma$ and
0.6$\sigma$), respectively.

\section{Timing analysis}
\label{sec:timing}

\begin{figure*}[ht]
  \begin{center}
    \leavevmode
    \epsscale{1.1}
    \plotone{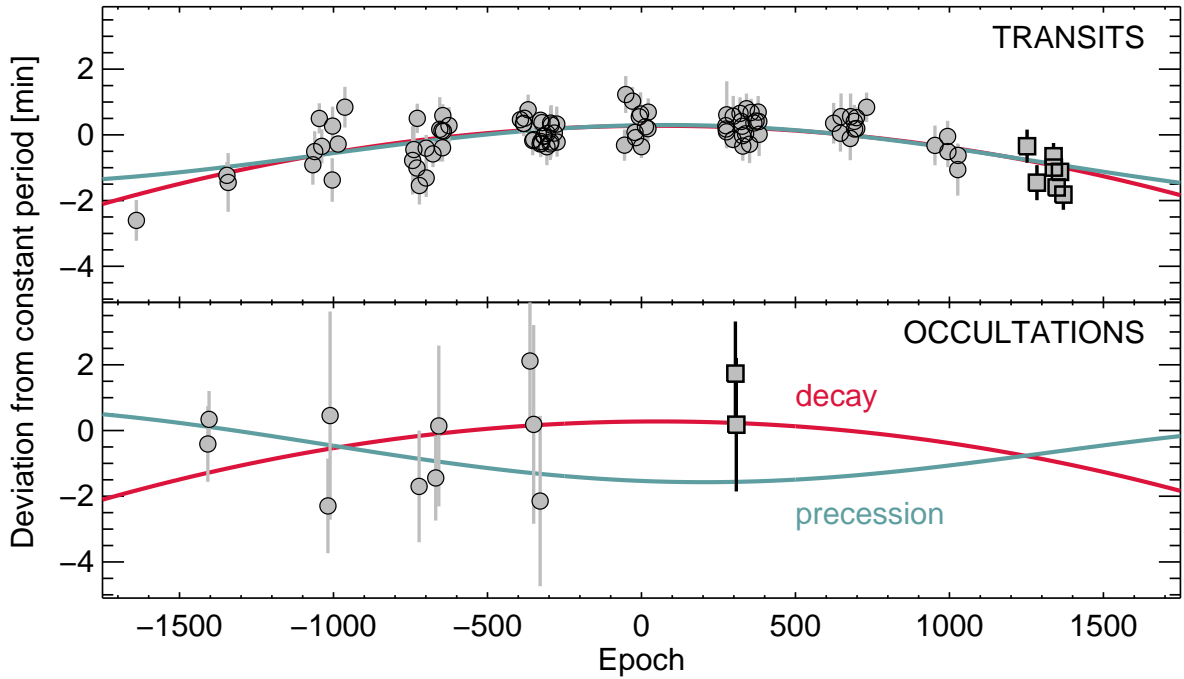}
  \end{center}
  \vspace{-0.2in}
  \caption{{\bf Timing residuals for WASP-12.} Each data point is the
    difference between an observed eclipse time and the prediction of
    the best-fit constant-period model. The top panel shows
    transit data and the bottom panel shows occultation data.  Circles
    are previously reported data, and squares are new data. The blue
    curves show the best-fit precession model, for which transit
    and occultation deviations are anticorrelated.  The red curves
    show the best-fit orbital decay model, in which the transit
    and occultation deviations are the same.
    \label{fig:times}}
\end{figure*}

Table~1 gathers together all of the times of transits
($t_{\rm t}$) and occultations ($t_{\rm o}$) used in our analysis. We
included all of the data we could find in the literature for which (i)
the analysis was based on observations of a single event, (ii) the
midpoint was allowed to be a completely free parameter, and (iii) the
time system is documented clearly.
The tabulated occultation times have not been corrected for the
light-travel time across the diameter of the orbit.  For the timing
analysis described below, the occultation times were corrected by
subtracting $2a/c = 22.9$~s.

We fitted three models to the timing data using the MCMC method.  The
first model assumes a circular orbit and a constant orbital period:
\begin{eqnarray}
  t_{\rm tra}(E) & = & t_0 + PE,\\
  t_{\rm occ}(E) & = & t_0 + \frac{P}{2} + PE,
\end{eqnarray}
where $E$ is the epoch number.
Figure~\ref{fig:times} displays the residuals with respect to this
model. The fit is poor, with $\chi^2_{\rm min} = 197.6$ and 111
degrees of freedom.  The transit residuals follow a negative parabolic
trend, indicating a negative period derivative.  Our new data---the
square points at the rightmost extreme of the plot---follow the trend
that had been established by \citet{Maciejewski+2016}.  Thus, we confirm the
finding of \citet{Maciejewski+2016} that the transit interval is
slowly shrinking.

Next we fitted a model that assumes a circular orbit and a constant period derivative:
\begin{eqnarray}
  t_{\rm tra}(E) & = & t_0 + PE + \frac{1}{2} \frac{dP}{dE} E^2,\\
  t_{\rm occ}(E) & = & t_0 + \frac{P}{2} + PE + \frac{1}{2} \frac{dP}{dE} E^2.
\end{eqnarray}
The red curves in Figure~\ref{fig:times} shows the best fit, which has $\chi^2_{\rm min}
= 118.5$ and 110 degrees of freedom.  Both the transit and occultation
data are compatible with the model.  The implied period derivative
is
\begin{equation}
  \frac{dP}{dt} = \frac{1}{P} \frac{dP}{dE} = -(9.3\pm 1.1)\times 10^{-10} = -29 \pm 3~{\rm ms~yr}^{-1}.
\end{equation}

In the third model, the orbit is slightly eccentric and
undergoing apsidal precession:
\begin{eqnarray}
  t_{\rm tra}(E) & = & t_0 + P_{\rm s}E - \frac{eP_{\rm a}}{\pi} \cos\omega,\\
  t_{\rm occ}(E) & = & t_0 + \frac{P_{\rm a}}{2} + P_{\rm s}E + \frac{eP_{\rm a}}{\pi} \cos\omega,
\end{eqnarray}
where $e$ is the eccentricity, $\omega$ is the argument of pericenter,
$P_{\rm a}$ is the anomalistic period and $P_{\rm s}$ is the sidereal period.
The argument of pericenter advances uniformly in time,
\begin{equation}
\omega(E) = \omega_0 + \frac{d\omega}{dE} E,
\end{equation}
and the two periods are related by
\begin{equation}
P_{\rm s} = P_{\rm a}\left(1 - \frac{d\omega/dE}{2\pi} \right).
\end{equation}
These expressions are based on
Eqn.~(15) of \citet{GimenezBastero1995}, in the limit of low eccentricity
and high inclination.
This model has 5 parameters: $t_0$, $P_{\rm s}$, $e$,
$\omega_0$, and $d\omega/dE$.

The blue curves in Figure~\ref{fig:times} show the best-fit
precession model. The main difference between the decay and precession
models is that apsidal precession produces anticorrelated transit and
occultation timing deviations, while the orbital decay model produces
deviations of the same sign. The precession fit has $\chi^2_{\rm
  min}=124.0$ and 108 degrees of freedom. The model achieves a
reasonable fit by adjusting the precession period to be longer than
the observing interval. In this way, the parabolic trend can be matched
by the downward-curving portion of a sinusoidal function. However,
there is tension between the need for enough downward curvature in the
transit deviations to fit the earliest data and a small enough upward
curvature in the occultation deviations to fit the most recent data.

The orbital decay model provides the best fit. It is better than the
precession model by $\Delta\chi^2 = 5.5$, despite the handicap of
having two fewer free parameters. The Akaike Information Criterion
(AIC) and Bayesian Information Criterion (BIC) are widely used
statistics to choose the most parsimonious model that fits the data:
\begin{eqnarray}
\alpha = {\rm AIC} &=& \chi^2 + 2k,\\
\beta = {\rm BIC} &=& \chi^2 + k\log n,
\end{eqnarray}
where $n$ is the number of data points and $k$ is the number of free
parameters. In this case, $n=113$, $k=3$ for decay, and $k=5$
for precession. The AIC favors the decay model by
$\Delta\alpha=9.46$, corresponding to a likelihood ratio of
$\exp(\Delta\alpha/2) = 113$. The BIC favors the orbital decay model
by $\Delta\beta=14.91$, corresponding to an approximate Bayes
factor of $\exp(\Delta\beta/2) = 1730$.

Table~2 gives the best-fit parameters for all three models. In
summary, a constant period has been firmly ruled out, and orbital
decay is statistically favored over apsidal precession as the best
explanation for the timing data. However, the statistical significance
of the preference for orbital decay is modest and depends on the
reliability of the quoted uncertainties for all of the timing data, which
come from different investigators using different methods. For
example, when the earliest data point is omitted, orbital decay is
still preferred but $\Delta\chi^2$ is reduced to 2.0.  For these
reasons, and out of general caution, we do not regard apsidal
precession as being definitively ruled out.  Further observations are
needed.

\section{Implications}
\label{sec:implications}

\subsection{Orbital decay}

To explore the implications of the best-fit models, we assume, for
the moment, that the orbital decay interpretation is correct.  Based on
the current decay rate, the period would shrink to zero in
\begin{equation}
\frac{P}{dP/dt} = 3.2~{\rm Myr}.
\end{equation}
The future lifetime of the planet is likely to be even shorter, because
the decay rate is expected to increase rapidly with decreasing period.

In the simplified ``constant phase lag'' model for
tidal evolution, the period derivative is
\begin{equation}
\label{eq:dpdt}
\frac{dP}{dt} = -\frac{27\pi}{2Q_\star}\left(\frac{M_{\rm p}}{M_\star}\right)\left(\frac{R_\star}{a}\right)^5,
\end{equation}
which we obtained by applying Kepler's third law to Eqn.~(20) of
\citet{GoldreichSoter1966}.  Here, $Q_\star$ is the ``modified quality
factor'' of the star's tidal oscillations (often designated elsewhere as
$Q_\star'$).  For the case of WASP-12, $M_{\rm p}/M_\star = 9.9\times
10^{-4}$ and $a/R_\star =3.097$ \citep{Chan+2011}, giving
\begin{equation}
Q_\star \approx 2\times 10^5.
\end{equation}

This value for $Q_\star$ is smaller than the typical range of
$10^{6-7}$ that has been inferred through ensemble analyses of binary
stars and star-planet systems \citep[see,
  e.g.][]{MeibomMathieu2005,Hansen2010,Penev+2012}. One exception is
\citet{Jackson+2008}, who found $Q_\star\sim 10^{5.5}$ based on the
period-eccentricity distribution of hot Jupiters. This is consistent
with our result.

Theoretically, the quality factor should depend on the orbital period,
perturbation strength, and internal structure of the star
\citep{Ogilvie2014}. Recently, \citet{EssickWeinberg2016} calculated
$Q_\star$ for hot Jupiters perturbing solar-type stars, based on the
nonlinear interactions and dissipation of tidally driven $g$-modes.
For the mass ratio and period of WASP-12, their Eqn.~(26) predicts
$Q_\star = 4\times 10^5$, close to the observed value. However, their
calculation pertained to stars with a radiative core and a convective
envelope, and it is not clear that WASP-12 belongs in this
category. With $T_{\rm eff}=6100$~K \citep{Torres+2012}, WASP-12 is
right on the borderline between stars with convective and radiative
envelopes. In fact, we wonder if this coincidence---lying right on the
Kraft break---could be related to the apparently rapid dissipation
rate. The star may have a convective core and a convective envelope,
separated by a radiative zone, perhaps leading to novel mechanisms for
wave dissipation.

\subsection{Apsidal precession}

Assuming instead that the apsidal precession model is correct, the
orbital eccentricity is $0.0021\pm 0.0005$. This is compatible with
the upper limit of 0.05 from observations of the spectroscopic orbit
\citep{Husnoo+2012}.  The observed precession rate is $\dot{\omega} =
26\pm 3$~deg~yr$^{-1}$, corresponding to a precession period of $14\pm
2$~years.

\citet{RagozzineWolf2009} showed that for systems resembling WASP-12,
the largest contribution to the theoretical apsidal precession rate is
from the planet's tidal deformation. The rate is proportional to the
planet's Love number $k_{\rm p}$, a dimensionless measure of the
degree of central concentration of the planet's density distribution.
Lower values of $k_{\rm p}$ correspond to more centrally concentrated
distributions, which are closer to the point-mass approximation and,
therefore, produce slower precession. Eqn.~(14) of
\citet{RagozzineWolf2009} can be rewritten for this case as
\begin{equation}
  \frac{d\omega}{dE} = 15\pi k_{\rm p} \left( \frac{M_\star}{M_{\rm p}} \right)
  \left( \frac{R_{\rm p}}{a} \right)^5.
\end{equation}
Using the measured precession rate and relevant parameters of WASP-12,
this equation gives $k_{\rm p} = 0.44\pm 0.10$. If this interpretation
is confirmed, it would be a unique constraint on an exoplanet's
interior structure, in addition to the usual measurements of mass,
radius, and mean density. For Jupiter, a value of $k_{\rm p}=0.59$ has
been inferred from its observed gravity moments \citep{Wahl+2016}.
Therefore the precession interpretation for WASP-12b suggests that its
density distribution has a similar degree of central concentration as
Jupiter, and perhaps somewhat higher.

\subsection{Prior probabilities}

It is worth contemplating the ``prior probability'' of each model. By
this, we mean the chance that the circumstances required by each model
would actually occur, independently of the goodness-of-fit to the
data. At face value, both models imply that we are observing WASP-12
at a special time, in violation of the ``temporal Copernican
principle'' articulated by \citet{Gott1993}. It is difficult, however,
to decide which model requires the greater coincidence.

Given the star's main-sequence age of $1700\pm 800$~Myr
\citep{Chan+2011}, the orbital decay model would have us believe we
are witnessing the last $\sim$0.2\% of the planet's life. If we
observed a single system at a random time, this would require a
one-in-500 coincidence. However, WASP-12 is not the only hot Jupiter
that we and others have been monitoring. There are about 10 other good
candidates with comparably low $a/R_\star$, increasing the odds of the
coincidence by an order of magnitude.

It is noteworthy that other investigators have argued on independent
grounds that WASP-12b is close to death.  \citet{Fossati+2010},
\citet{Haswell2012}, and \citet{Nichols+2015} have presented
near-ultraviolet transit spectroscopy consistent with an extended and
escaping exosphere.  The resulting mass loss process has been studied
theoretically by \citet{Li+2010}, \citet{Lai+2010}, and
\citet{Bisikalo+2013}. Most recently, \citet{Jackson+2017} developed a
new theory for Roche lobe overflow and identified WASP-12 as a likely
case of rapid mass loss.

It is also possible that orbital decay occurs in fits and starts,
because of strong and erratic variations in the dissipation rate with
the forcing period \citep[see,
  e.g.,][]{OgilvieLin2007,BarkerOgilvie2010}. Thus, the planet may be
experiencing a brief interval of rapid decay. This does not eliminate
the requirement for a coincidence, because one would expect to
discover the system in one of the more prolonged states of slow
dissipation. However, it does mean that the planet's future lifetime
may be longer than the current value of $P/\dot{P}$.

As for apsidal precession, the trouble is the very short expected
timescale for tidal orbital circularization. This process is thought
to be dominated by dissipation within the planet, rather than the
star.  Eqn.~(25) of \citet{GoldreichSoter1966}, relevant to this case,
can be rewritten
\begin{equation}
\tau_e = \frac{e}{|de/dt|} =
\frac{2Q_{\rm p}}{63\pi} \left(\frac{M_{\rm p}}{M_\star}\right)
\left( \frac{a}{R_{\rm p}} \right)^5 P_{\rm orb}.
\end{equation}
For WASP-12, this gives $\tau_e \sim 0.5$~Myr, assuming $Q_{\rm p}\sim
10^6$.  At this rate, even 4~Myr of tidal evolution would reduce the
eccentricity below $10^{-3}$.  Of course, the planetary quality
factor $Q_p$ could be larger than the standard value of $10^6$, or
the tidal model leading to the preceding equation could be a gross
misrepresentation of the actual circularization process.

There may also be some process that continually excites the
eccentricity. One possibility is gravitational forcing by another
planet, although no other nearby planets are known in the WASP-12
system \citep{Knutson+2014}.  An intriguing possibility is eccentricity excitation by the
gravitational perturbations from the star's convective eddies. In this
scenario, proposed by \citet{Phinney1992} to explain the small but
nonzero eccentricities of pulsars orbiting white dwarfs, the system
reaches a state of equipartition between the energy of eccentricity
oscillations (epicyclic motion) and the kinetic energy of turbulent
convection. To our knowledge, this theory has only been developed for
post-main-sequence stars \citep[see,
  e.g.][]{VerbuntPhinney1995,Rafikov2016}. It is not obvious that this
theory would apply to WASP-12 and be compatible with $e\sim10^{-3}$.

Should further theoretical investigations reveal that this mechanism
(or any other) could naturally maintain the orbital eccentricity at
the level of $10^{-3}$, then the apsidal precession model would
require no special coincidence. Neither would it require unique
circumstances; it is possible that eccentricities of this order could
exist in other hot Jupiter systems and have remained undetected. Thus,
the identification of a natural eccentricity-excitation mechanism
would swing the prior-probability balance in the direction of apsidal
precession.

\subsection{Other possible explanations}

To this point, we have presented orbital decay and apsidal precession
as the only possible reasons for an apparent period decrease. Another
possibility is that the star is accelerating toward Earth, due to the
force from stellar companions or wide-orbiting planets.  This would
produce a negative apparent period derivative of $\dot{v}_{\rm r}P/c$,
where $v_{\rm r}$ is the radial velocity. Based on long-term Doppler
monitoring, \citet{Knutson+2014} placed an upper limit for WASP-12 of
$|\dot{v}_{\rm r}| < 0.019$~m~s$^{-1}$~day$^{-1}$ (2$\sigma$), which,
in turn, limits the apparent $\dot{P}$ to be smaller than $7\times
10^{-11}$. This is an order of magnitude too small to be responsible
for the observed period derivative. Of course, none of these phenomena
are mutually exclusive. The system may be experiencing a combination
of orbital decay, apsidal precession, and radial acceleration,
although joint modeling of these effects is not productive with the
current data.

\citet{Rafikov2009} described two other phenomena that cause changes
in the apparent period of a transiting planet. The first is the
Shklovskii effect, wherein the star's proper motion leads to a
changing radial velocity and a nonzero second derivative of the
light-travel time. This is already ruled out by Doppler observations
of the radial acceleration. For completeness, though, we note that the
observed distance $d$ and proper motion $\mu$ imply a period
derivative of $P\mu^2 d/c \sim 6\times 10^{-15}$, too small to explain
the data. The second phenomenon, also dependent on proper motion, is
the apparent apsidal precession caused by our changing viewing
angle. The resulting period derivative is of order
$\sim$$(P\mu)^2/2\pi$, which in this case is $\sim$$10^{-21}$, too
small by many orders of magnitude.

\section{Future prospects}
\label{sec:future}

\begin{figure*}[ht!]
  \begin{center}
    \leavevmode
    \epsscale{1.1}
    \plotone{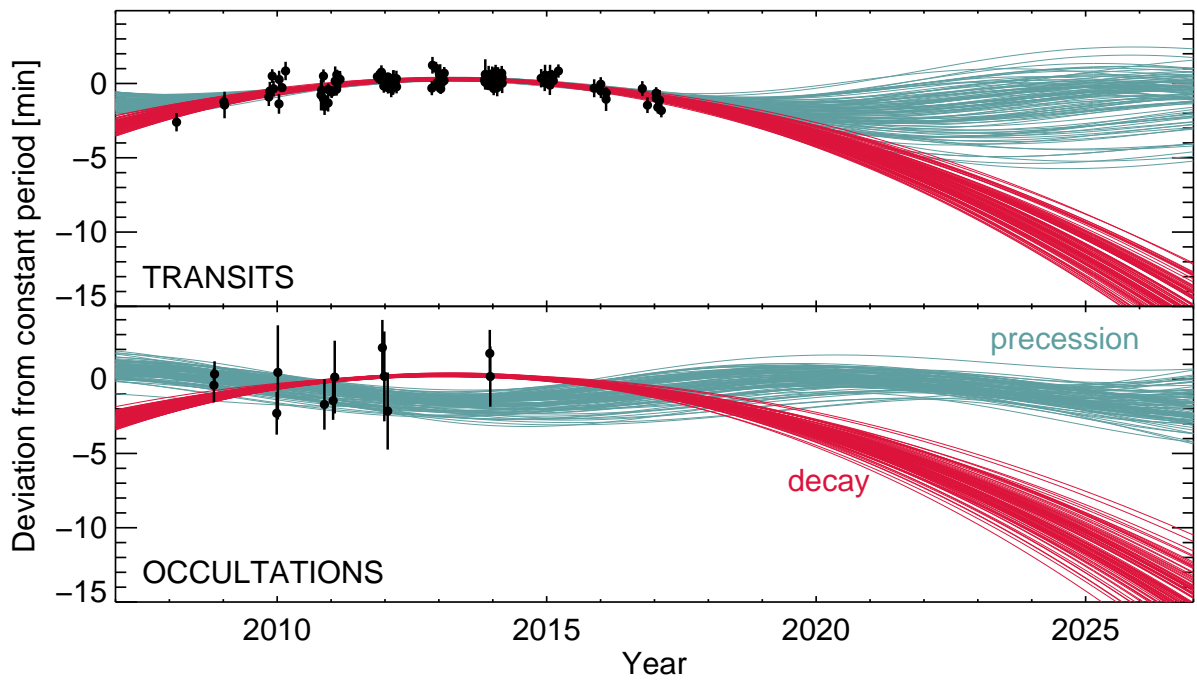}
  \end{center}
  \vspace{-0.2in}
  \caption{{\bf Possible futures for WASP-12.} For each of the two
    models, we randomly drew 100 parameter sets from our Markov
    chains. Shown here are the extrapolations of those models to
    future times.
    \label{fig:future}}
\end{figure*}

With WASP-12, we are fortunate that both possibilities---orbital decay
and apsidal precession---lead to interesting outcomes.  It will soon
be possible to measure the tidal dissipation rate of a star, or the
tidal deformability of an exoplanet, either of which would be a unique
achievement.  To help understand the requirements for a definitive
verdict, Figure~\ref{fig:future} shows the future projections of a
sample of 100 models that provide satisfactory fits to the data, drawn
randomly from our converged Markov chains.

For the transits, the two families of models become separated by a few
minutes by 2021-22.  The occultation models diverge earlier, and are
separated by a few minutes in 2019-20.  Thus, while continued transit
timing is important, the most rapid resolution would probably come
from observing occultations a few years from now.  In principle,
transit duration variations (TDV) would also help to distinguish
between the two models, but the expected amplitude is
\citep{PalKocsis2008}
\begin{equation}
{\rm TDV} \sim \frac{P}{2\pi} \left( \frac{R_\star}{a}\right) e\cos\omega \sim 10~{\rm sec},
\end{equation}
which will be difficult to detect.

In this paper, we have focused on the timing anomalies of WASP-12. This
system has other remarkable features we have not even discussed.  The
star's equator is likely to be misaligned with the orbital plane
\citep{Schlaufman2010,Albrecht+2012}. The star is also part of a
hierarchical three-body system, with a tight pair of M dwarfs orbiting
the planet-hosting star at a distance of about 265~AU
\citep{Bechter+2014}. Detailed modeling of the star's interior
structure and and tidal evolution is warranted, as are continued
observations of transits and occultations.

\acknowledgements We are very grateful to Allyson Bieryla, David
Latham, and Emilio Falco for their assistance with the FLWO
observations. We thank Nevin Weinberg, Jeremy Goodman, Kaloyen Penev,
David Oort Alonso, Heather Knutson, Dong Lai, and the CfA Exoplanet
Pizza group for helpful discussions. We also appreciate the anonymous
reviewer's prompt and careful report. Work by K.C.P.\ was supported by
the MIT Undergraduate Research Opportunities Program and the Paul
E.\ Gray Fund.

\vskip 0.1in
\small
{\it Note added in proof.} D.\ Lai has reminded us of another possible
reason for a cyclic variation in the period:
the \citet{Applegate1992} effect, in which a star's quadrupole moment
varies over a magnetic activity cycle.  For WASP-12,
\citet{WatsonMarsh2010} estimated that this effect could produce
timing deviations of 4-40~s depending on the cycle duration.  The
transit and occultation deviations would have the same sign,
allowing this effect to be distinguished from apsidal precession.

\clearpage

\begin{deluxetable}{ccccl}
  \label{tbl:times}
\tabletypesize{\scriptsize}
\tablewidth{0pt}
\tablecaption{Transit and occultation times.}
\tablehead{
  \colhead{Type of} &
  \colhead{Midpoint} &
  \colhead{Uncertainty} &
  \colhead{Epoch} \\
  \colhead{event} &
  \colhead{(BJD$_{\rm TDB}$)} &
  \colhead{(days)} &
  \colhead{number}
}
\renewcommand{\arraystretch}{0.8}
\startdata
tra & 2454515.52496 &  0.00043 &  -1640  & \citet{Hebb+2009}\tablenotemark{a}\\
occ & 2454769.28131 &  0.00080 &  -1408  & \citet{Campo+2011}\\
occ & 2454773.64751 &  0.00060 &  -1404  & \citet{Campo+2011}                                       \\
tra & 2454836.40340 &  0.00028 &  -1346  & \citet{Copperwheat+2013}                                  \\
tra & 2454840.76893 &  0.00062 &  -1342  & \citet{Chan+2011}                                         \\
tra & 2455140.90981 &  0.00042 &  -1067  & \citet{Collins+2017}                                      \\
tra & 2455147.45861 &  0.00043 &  -1061  & \citet{Maciejewski+2013}                                  \\
tra & 2455163.83061 &  0.00032 &  -1046  & \citet{Collins+2017}                                      \\
tra & 2455172.56138 &  0.00036 &  -1038  & \citet{Chan+2011}                                         \\
occ & 2455194.93381 &  0.00100 &  -1018  & \citet{Croll+2015}\\
occ & 2455202.57566 &  0.00220 &  -1011  & \citet{Fohring+2013}\\
tra & 2455209.66895 &  0.00046 &  -1004  & \citet{Collins+2017}                                      \\
tra & 2455210.76151 &  0.00041 &  -1003  & \citet{Collins+2017}                                      \\
tra & 2455230.40669 &  0.00019 &   -985  & \citet{Maciejewski+2013}\\
tra & 2455254.41871 &  0.00043 &   -963  & \citet{Maciejewski+2013}\\
tra & 2455494.52999 &  0.00072 &   -743  & \citet{Maciejewski+2013}                                  \\
tra & 2455498.89590 &  0.00079 &   -739  & \citet{Sada+2012}                                         \\
tra & 2455509.80971 &  0.00037 &   -729  & \citet{Collins+2017}                                      \\
tra & 2455510.90218 &  0.00031 &   -728  & \citet{Collins+2017}                                      \\
occ & 2455517.99455 &  0.00118 &   -722  & \citet{Deming+2015}\tablenotemark{b}\\
tra & 2455518.54070 &  0.00040 &   -721  & \citet{Cowan+2012}                                        \\
tra & 2455542.55210 &  0.00040 &   -699  & \citet{Cowan+2012}                                        \\
tra & 2455542.55273 &  0.00028 &   -699  & \citet{Maciejewski+2013}                                  \\
tra & 2455566.56385 &  0.00028 &   -677  & \citet{Maciejewski+2013}                                  \\
occ & 2455576.93141 &  0.00090 &   -668  & \citet{Croll+2015}\\
occ & 2455587.84671 &  0.00170 &   -658  & \citet{Croll+2015}\\
tra & 2455590.57561 &  0.00068 &   -655  & \citet{Maciejewski+2013}                                  \\
tra & 2455598.21552 &  0.00035 &   -648  & \citet{Maciejewski+2013}                                  \\
tra & 2455600.39800 &  0.00029 &   -646  & \citet{Maciejewski+2013}                                  \\
tra & 2455601.49010 &  0.00024 &   -645  & \citet{Maciejewski+2013}                                  \\
tra & 2455603.67261 &  0.00029 &   -643  & \citet{Collins+2017}                                      \\
tra & 2455623.31829 &  0.00039 &   -625  & \citet{Maciejewski+2013}                                  \\
tra & 2455876.52786 &  0.00027 &   -393  & \citet{Maciejewski+2013}                                  \\
tra & 2455887.44198 &  0.00021 &   -383  & \citet{Maciejewski+2013}                                  \\
tra & 2455888.53340 &  0.00027 &   -382  & \citet{Maciejewski+2013}                                  \\
tra & 2455890.71635 &  0.00024 &   -380  & \citet{Maciejewski+2013}                                  \\
tra & 2455903.81357 &  0.00032 &   -368  & \citet{Collins+2017}                                      \\
occ & 2455910.90841 &  0.00130 &   -362  & \citet{Crossfield+2012}\\
tra & 2455920.18422 &  0.00031 &   -353  & \citet{Maciejewski+2013}                                  \\
tra & 2455923.45850 &  0.00022 &   -350  & \citet{Maciejewski+2013}                                  \\
tra & 2455924.00411 &  0.00210 &   -350  & \citet{Croll+2015}\\
tra & 2455946.37823 &  0.00018 &   -329  & \citet{Maciejewski+2013}                                  \\
occ & 2455946.92231 &  0.00180 &   -329  & \citet{Croll+2015}\\
tra & 2455947.47015 &  0.00017 &   -328  & \citet{Maciejewski+2013}                                  \\
tra & 2455948.56112 &  0.00033 &   -327  & \citet{Maciejewski+2013}                                  \\
tra & 2455951.83534 &  0.00011 &   -324  & \citet{Stevenson+2014}                                    \\
tra & 2455952.92720 &  0.00010 &   -323  & \citet{Stevenson+2014}                                    \\
tra & 2455959.47543 &  0.00017 &   -317  & \citet{Maciejewski+2013}                                  \\
tra & 2455960.56686 &  0.00032 &   -316  & \citet{Maciejewski+2013}                                  \\
tra & 2455970.38941 &  0.00039 &   -307  & \citet{Maciejewski+2013}                                  \\
tra & 2455971.48111 &  0.00035 &   -306  & \citet{Maciejewski+2013}                                  \\
tra & 2455982.39509 &  0.00034 &   -296  & \citet{Maciejewski+2013}                                  \\
tra & 2455983.48695 &  0.00035 &   -295  & \citet{Maciejewski+2013}                                  \\
tra & 2455984.57797 &  0.00032 &   -294  & \citet{Collins+2017}                                      \\
tra & 2455985.66975 &  0.00042 &   -293  & \citet{Collins+2017}                                      \\
tra & 2455996.58378 &  0.00037 &   -283  & \citet{Collins+2017}                                      \\
tra & 2456005.31533 &  0.00037 &   -275  & \citet{Maciejewski+2013}                                  \\
tra & 2456006.40637 &  0.00031 &   -274  & \citet{Maciejewski+2013}                                  \\
tra & 2456245.42729 &  0.00033 &    -55  & \citet{Maciejewski+2016}                                  \\
tra & 2456249.79404 &  0.00039 &    -51  & \citet{Collins+2017}                                      \\
tra & 2456273.80514 &  0.00030 &    -29  & \citet{Collins+2017}                                      \\
tra & 2456282.53584 &  0.00030 &    -21  & \citet{Maciejewski+2016}                                  \\
tra & 2456284.71857 &  0.00030 &    -19  & \citet{Collins+2017}                                      \\
tra & 2456297.81605 &  0.00030 &     -7  & \citet{Collins+2017}                                      \\
tra & 2456302.18179 &  0.00046 &     -3  & \citet{Maciejewski+2016}                                  \\
tra & 2456305.45536 &  0.00024 &      0  & \citet{Maciejewski+2016}                                  \\
tra & 2456319.64424 &  0.00038 &     13  & \citet{Collins+2017}                                      \\
tra & 2456328.37556 &  0.00027 &     21  & \citet{Maciejewski+2016}                                  \\
tra & 2456329.46733 &  0.00029 &     22  & \citet{Maciejewski+2016}                                  \\
tra & 2456604.50489 &  0.00021 &    274  & \citet{Maciejewski+2016}                                  \\
tra & 2456605.59624 &  0.00030 &    275  & \citet{Maciejewski+2016}                                  \\
tra & 2456606.68760 &  0.00033 &    276  & \citet{Maciejewski+2016}                                  \\
tra & 2456607.77938 &  0.00071 &    277  & \citet{Collins+2017}                                      \\
tra & 2456629.60726 &  0.00019 &    297  & \citet{Maciejewski+2016}                                  \\
tra & 2456630.69917 &  0.00043 &    298  & \citet{Maciejewski+2016}                                  \\
occ & 2456638.88530 &  0.00110 &    305  & this work\\
occ & 2456642.15848 &  0.00141 &    308  & this work\\
tra & 2456654.71047 &  0.00034 &    320  & \citet{Collins+2017}                                      \\
tra & 2456659.07598 &  0.00034 &    324  & \citet{Kreidberg+2015}                                    \\
tra & 2456662.35014 &  0.00019 &    327  & \citet{Maciejewski+2016}                                  \\
tra & 2456663.44136 &  0.00019 &    328  & \citet{Maciejewski+2016}                                  \\
tra & 2456664.53256 &  0.00031 &    329  & \citet{Maciejewski+2016}                                  \\
tra & 2456674.35560 &  0.00028 &    338  & \citet{Kreidberg+2015}                                    \\
tra & 2456677.63039 &  0.00032 &    341  & \citet{Collins+2017}                                      \\
tra & 2456688.54384 &  0.00040 &    351  & \citet{Maciejewski+2016}                                  \\
tra & 2456694.00161 &  0.00029 &    356  & \citet{Kreidberg+2015}                                    \\
tra & 2456703.82417 &  0.00029 &    365  & \citet{Kreidberg+2015}                                    \\
tra & 2456711.46415 &  0.00025 &    372  & \citet{Maciejewski+2016}                                  \\
tra & 2456719.10428 &  0.00034 &    379  & \citet{Kreidberg+2015}                                    \\
tra & 2456721.28692 &  0.00034 &    381  & \citet{Kreidberg+2015}                                    \\
tra & 2456722.37807 &  0.00046 &    382  & \citet{Maciejewski+2016}                                  \\
tra & 2456986.50195 &  0.00043 &    624  & \citet{Maciejewski+2016}                                  \\
tra & 2457010.51298 &  0.00039 &    646  & \citet{Maciejewski+2016}                                  \\
tra & 2457012.69617 &  0.00049 &    648  & \citet{Collins+2017}                                      \\
tra & 2457045.43831 &  0.00046 &    678  & \citet{Maciejewski+2016}                                  \\
tra & 2457046.53019 &  0.00049 &    679  & \citet{Maciejewski+2016}                                  \\
tra & 2457059.62713 &  0.00035 &    691  & \citet{Collins+2017}                                      \\
tra & 2457060.71839 &  0.00036 &    692  & \citet{Collins+2017}                                      \\
tra & 2457067.26715 &  0.00022 &    698  & \citet{Maciejewski+2016}                                  \\
tra & 2457068.35834 &  0.00020 &    699  & \citet{Maciejewski+2016}                                  \\
tra & 2457103.28423 &  0.00031 &    731  & \citet{Maciejewski+2016}                                  \\
tra & 2457345.57867 &  0.00042 &    953  & \citet{Maciejewski+2016}                                  \\
tra & 2457390.32708 &  0.00033 &    994  & \citet{Maciejewski+2016}                                  \\
tra & 2457391.41818 &  0.00033 &    995  & \citet{Maciejewski+2016}                                  \\
tra & 2457426.34324 &  0.00055 &   1027  & \citet{Maciejewski+2016}                                  \\
tra & 2457427.43496 &  0.00023 &   1028  & \citet{Maciejewski+2016}                                 \\
tra & 2457671.91324 &  0.00035 &   1252  & this work\\
tra & 2457706.83791 &  0.00037 &   1284  & this work\\
tra & 2457765.77515 &  0.00028 &   1338  & this work\\
tra & 2457766.86633 &  0.00039 &   1339  & this work                                      \\
tra & 2457776.68869 &  0.00029 &   1348  & this work                                     \\
tra & 2457788.69464 &  0.00048 &   1359  & this work                                       \\
tra & 2457800.69978 &  0.00032 &   1370  & this work
\enddata
\tablenotetext{a}{Refers to the light curve obtained by
  \citet{Hebb+2009} with the 2m Liverpool telescope, as analyzed by
  \citet{Maciejewski+2013}.}
\tablenotetext{b}{Re-analyzed in this work.}
\end{deluxetable}

\renewcommand{\arraystretch}{1.0}
\begin{deluxetable}{lc}
      \label{tbl:parameters}
\tabletypesize{\small}
\tablewidth{0pt}
\tablecaption{Best-fit model parameters.}
\tablehead{
  \colhead{Parameter} &
  \colhead{Value~(Unc.)\tablenotemark{a}}
}
\startdata
~~~~~~{\it Constant period} &  \\
Reference epoch, $t_0$~[BJD$_{\rm TBD}$] & 2456305.455609(28)\\
Period, $P$~[days]                      & 1.091420025(47)\\
~~~~~~{\it Orbital decay} &  \\
Reference epoch, $t_0$~[BJD$_{\rm TBD}$] &   2456305.455790(35)\\
Period at reference epoch, $P$~[days]   &  1.091420078(47) \\
$dP/dE$~[days]                          &  $-1.02(11) \times 10^{-9}$\\
~~~~~~{\it Apsidal precession} &  \\
Reference epoch, $t_0$~[BJD$_{\rm TBD}$] &   2456305.45509(15)\\
Sidereal period, $P_{\rm sid}$~[days]    &         1.09141993(15)\\
Eccentricity, $e$                      &         0.00208(47)\\
A.O.P.\ at reference epoch, $\omega_0$~[rad] &         2.92(19)\\
Precession rate, $d\omega/dE$~[rad~epoch$^{-1}$] &   0.00133(18)
\enddata
\tablenotetext{a}{The numbers in parenthesis give the 1$\sigma$ uncertainty in the final
  two digits.}
\end{deluxetable}

\clearpage

\bibliographystyle{yahapj}
\bibliography{references}

\end{document}